\documentclass[10pt]{article}%
\usepackage{amssymb}
\usepackage{graphicx}
\usepackage{amsmath}
\usepackage{amsthm}
\usepackage{amsfonts}%
\newtheorem*{thm1}{Theorem 1}
\newtheorem*{thm2}{Theorem 2}
\begin{document}

\title{Small-Signal Amplification of Period-Doubling Bifurcations in Smooth Iterated Maps}
\author{Xiaopeng Zhao,$^{1}$\thanks{Corresponding author. Email: xzhao@duke.edu}%
~~David G. Schaeffer,$^{2}$ Carolyn M. Berger,$^{3}$ and Daniel J.
Gauthier,$^{1,3}$\\Department of $^{1}$Biomedical Engineering, $^{2}$Mathematics, and $^{3}
$Physics \\and Center for Nonlinear and Complex Systems \\Duke University, Durham, NC 27708}
\date{\today}
\maketitle

\begin{abstract}
Various authors have shown that, near the onset of a period-doubling
bifurcation, small perturbations in the control parameter may result in much
larger disturbances in the response of the dynamical system. Such
amplification of small signals can be measured by a gain defined as the
magnitude of the disturbance in the response divided by the perturbation
amplitude. In this paper, the perturbed response is studied using normal forms
based on the most general assumptions of iterated maps. Such an analysis
provides a theoretical footing for previous experimental and numerical
observations, such as the failure of linear analysis and the saturation of the
gain. Qualitative as well as quantitative features of the gain are exhibited
using selected models of cardiac dynamics.

\end{abstract}

\section{Introduction}

We study a dynamical system described by a one-parameter family of iterated
maps
\begin{equation}
x_{n+1}=f\left(  x_{n};\mu\right)  , \label{eqn:generalMap}%
\end{equation}
where the state $x$ may be one- or multi-dimensional. Suppose that a fixed
point of the map experiences a supercritical period-doubling bifurcation when
the control parameter $\mu$ passes through a critical value $\mu_{\text{bif}}%
$. Near onset of the bifurcation, the system is sensitive to small alternating
perturbations applied to $\mu$ [1-14]. Under such alternating perturbations,
the governing map becomes%
\begin{equation}
x_{n+1}=f\left(  x_{n};\mu+\left(  -1\right)  ^{n}\delta\right)  ,
\label{eqn:perturbedMap}%
\end{equation}
where $\delta$ is the perturbation amplitude. Consequently, the steady-state
solution of the system alternates between two states, $x_{\text{even}}$ and
$x_{\text{odd}}$, in the following manner:%
\begin{align}
x_{\text{even}}  &  =f\left(  x_{\text{odd}};\mu-\delta\right) \\
x_{\text{odd}}  &  =f\left(  x_{\text{even}};\mu+\delta\right)  .
\end{align}
The magnitude of deviation between $x_{\text{even}}$ and $x_{\text{odd}}$
could be many times larger than the perturbation amplitude $\delta$, an effect
which is known as pre-bifurcation amplification [7-9]. To characterize this
amplification, one may define a gain using the $i$th component of $x$ as
follows%
\begin{equation}
\text{$\Gamma$}\equiv\left\vert \frac{x_{\text{even}}^{\left(  i\right)
}-x_{\text{odd}}^{\left(  i\right)  }}{2\,\delta}\right\vert .
\label{eqn:gaindef}%
\end{equation}
In this work, we explore qualitative changes in the gain $\Gamma$ under
changes in the control parameter $\mu$ and perturbation amplitude $\delta$.

It has been shown that pre-bifurcation amplification is a universal phenomenon
existing in discrete [1-6] as well as continuous [7-10] dynamical systems.
Many authors utilized one-dimensional maps to illustrate the scaling of
parameters. Such one-dimensional maps either take the most general form as in
Heldstab \emph{et al.} \cite{heldstab83} or take the normal form as in Bryant
and Wiesenfeld \cite{bryant86}. Since the normal form in \cite{bryant86} was
not related to the full-dimensional underlying model, it is not clear how the
scaling of parameters is connected to physical properties of the system.

Although the fundamental scaling law behind prebifurcation amplifiction has
been shown in the literature, explicit solutions of the gain $\Gamma$ are not
known for arbitrarily-given iterated maps. In this paper, we aim to derive an
analytical formula of the gain based on multi-dimensional iterated maps,
requiring only minimum assumptions. Using numerical simulations as well as
experimental work \cite{bryant86,vohra91,vohra95}, a few researchers have
shown that the correct scaling law must resort to a nonlinear analysis. We
illustrate, for the first time, the failure of the linear analysis through an
order analysis, which reveals that the range of validity for the linear
analysis is extremely limited and dramatically shrinks to zero as the
bifurcation point is approached. We also explain the saturation of
amplification previously observed in \cite{kravtsov03,surovyatkina04} using
the developed scaling law.

Another objective of the current work is to examine period-doubling
bifurcations in cardiac dynamics, \emph{i.e.}, alternans in cardiac tissues,
which are characterized by the long-short beat-to-beat alternation in the
duration of cardiac action potentials [15-18]. Cardiac alternans have been
recognized as a possible initiator of fatal arrhythmia, which is the number
one cause of death in the United States \cite{AhA2004}. We apply the
theoretical results to two cardiac mappings to help gain an understanding of
the origin and control of instabilities in cardiac tissues.

The paper is organized as follows. In Section 2, we compute the response of
alternating perturbations using linear analysis of the map. A nonlinear
analysis based on higher-order approximation of the map is then performed in
Section 3. A summary and discussion is presented in Section 4.

\section{Linear Analysis}

At $\mu=\mu_{\text{bif}}$, denote the fixed point of the map
(\ref{eqn:generalMap}) by $x_{\text{bif}}$. Then the Jacobian $f_{x}\left(
x_{\text{bif}};\mu_{\text{bif}}\right)  $ has one eigenvalue equal to $-1$
with the corresponding right and left eigenvectors denoted by $\phi_{1}$ and
$\psi_{1}$, respectively. Near onset of the period-doubling bifurcation, the
response of the perturbed map (\ref{eqn:perturbedMap}) exhibits
pre-bifurcation amplification. The following theorem computes the
amplification gain based on linearization of the perturbed map.

\begin{thm1}
In linear approximation, as $\mu\rightarrow\mu_{\text{bif}}$, the gain defined
in Eq. (\ref{eqn:gaindef}) diverges as%
\begin{equation}
\text{$\Gamma\sim$}\left\vert \frac{k}{\mu-\mu_{\text{bif}}}\right\vert ,
\end{equation}
where the constant $k$ is determined by data from the mapping. Specifically,%
\begin{equation}
k=\frac{\psi_{1}^{T}\cdot f_{\mu}\left(  x_{\text{bif}};\mu_{\text{bif}%
}\right)  }{\left(  \psi_{1}^{T}\cdot L_{1}\cdot\phi_{1}\right)  }\phi
_{1}^{\left(  i\right)  },
\end{equation}
where%
\begin{equation}
L_{1}=f_{xx}\left(  x_{\text{bif}};\mu_{\text{bif}}\right)  \cdot
\frac{\partial x_{\ast}}{\partial\mu}\left(  \mu_{\text{bif}}\right)
+f_{x\mu}\left(  x_{\text{bif}};\mu_{\text{bif}}\right)  ,
\end{equation}
and $x_{\ast}\left(  \mu\right)  $ is a fixed point of map
(\ref{eqn:generalMap}).
\end{thm1}

In Theorem 1, $\phi_{1}^{\left(  i\right)  }$ represents the $i$th component
of $\phi_{1}$. For one-dimensional maps, the eigenvectors are $\phi_{1}%
=\psi_{1}=1$ and thus%
\begin{equation}
k=\frac{f_{\mu}\left(  x_{\text{bif}};\mu_{\text{bif}}\right)  }{f_{xx}\left(
x_{\text{bif}};\mu_{\text{bif}}\right)  \cdot\frac{\partial x_{\ast}}%
{\partial\mu}\left(  \mu_{\text{bif}}\right)  +f_{x\mu}\left(  x_{\text{bif}%
};\mu_{\text{bif}}\right)  }.
\end{equation}
Note, here, subscripts indicate partial derivatives of the function $f$. This
theorem indicates that the magnitude of the gain depends on the properties of
the system at the bifurcation point. Moreover, one can make two striking
observations regarding the linear approximation: i) the gain does not depend
on the amplitude $\delta$ and ii) the gain approaches $\infty$ as
$\mu\rightarrow\mu_{\text{bif}}$. As shown at the end of this section, both
observations compare poorly with numerical simulations. To rectify the
discrepancy, we are led to the higher-order nonlinear analysis of the gain in
Section 3.

\subsection{Proof of Theorem 1}

We prove the theorem by a perturbation-theory calculation. We assume the map
(\ref{eqn:generalMap}) is $N$-dimensional. Denote the fixed point of map
(\ref{eqn:generalMap}) as $x_{\ast}\left(  \mu\right)  $, so that%
\begin{equation}
x_{\ast}\left(  \mu\right)  =f\left(  x_{\ast}\left(  \mu\right)  ;\mu\right)
. \label{eqn:md_fix}%
\end{equation}
Let us assume the fixed point is stable when $\mu$ is greater than but close
to $\mu_{\text{bif}}$ (otherwise we could introduce a new control parameter as
$\bar{\mu}=-\mu$). We then study the response of the perturbed map
(\ref{eqn:perturbedMap}). Under small perturbation, the response of the
perturbed system is a small deviation (denoted by $d_{n}$) from the
unperturbed fixed point. Let us write $x_{n}=x_{\ast}+d_{n}$, which upon
substitution into Eq. (\ref{eqn:perturbedMap}), yields%
\begin{equation}
d_{n+1}=L\cdot d_{n}+\left(  -1\right)  ^{n}f_{\mu}\left(  x_{\ast}%
;\mu\right)  \delta, \label{eqn:linear_dn}%
\end{equation}
where%
\begin{equation}
L=f_{x}\left(  x_{\ast};\mu\right)  .
\end{equation}
To account for the alternating feature of the perturbation, we decompose the
steady state of $d_{n}$ into an oscillating component $\left(  -1\right)
^{n}v$ and a mean value $m$: \emph{i.e.}, $d_{n}=\left(  -1\right)  ^{n}v+m$.
Substituting $d_{n}$ into Eq. (\ref{eqn:linear_dn}) yields%
\begin{align}
m  &  =L\cdot m\\
v  &  =-L\cdot v-f_{\mu}\left(  x;\mu\right)  \delta. \label{eqn:md_sn1}%
\end{align}
It follows that $m=0$ in this linear approximation.

In the following, we find a solution for $v$. Because $\mu$ is close to
$\mu_{\text{bif}}$, $L$ can be approximated as
\begin{equation}
L=L_{0}+\left(  \mu-\mu_{\text{bif}}\right)  L_{1}, \label{eqn:md_L}%
\end{equation}
where%
\begin{align}
L_{0}  &  =f_{x}\left(  x_{\text{bif}};\mu_{\text{bif}}\right)
\label{eqn:md_L0}\\
L_{1}  &  =f_{xx}\left(  x_{\text{bif}};\mu_{\text{bif}}\right)  \cdot
\frac{\partial x_{\ast}}{\partial\mu}\left(  \mu_{\text{bif}}\right)
+f_{x\mu}\left(  x_{\text{bif}};\mu_{\text{bif}}\right)  . \label{eqn:md_L1}%
\end{align}
Implicit differentiation of Eq. (\ref{eqn:md_fix}) yields
\begin{equation}
\frac{\partial x_{\ast}}{\partial\mu}\left(  \mu_{\text{bif}}\right)  =\left(
\mathbf{I}-f_{x}\left(  x_{\text{bif}};\mu_{\text{bif}}\right)  \right)
^{-1}\cdot f_{\mu}\left(  x_{\text{bif}};\mu_{\text{bif}}\right)  ,
\label{eqn:md_xmu}%
\end{equation}
where $\mathbf{I}$ stands for the identity matrix. Denote the eigenvalues of
$L_{0}$ by $-1=\lambda_{1}<\lambda_{2}\leq\ldots\leq\lambda_{N}<1$ and denote
the corresponding right and left eigenvectors by $\phi_{1},\phi_{2}%
,\ldots,\phi_{N}$ and $\psi_{1}^{T},\psi_{2}^{T},\ldots,\psi_{N}^{T}$,
respectively. Assume the eigenvectors are normalized such that $\psi_{i}%
^{T}\cdot\phi_{j}=1$ if $i=j$ and $\psi_{i}^{T}\cdot\phi_{j}=0$ if $i\neq j$.
We can decompose $v$ as%
\begin{equation}
v=\sum_{i=1}^{N}s_{i}\phi_{i}. \label{eqn:v_decompose}%
\end{equation}
Substituting Eq. (\ref{eqn:v_decompose}) into Eq. (\ref{eqn:md_sn1}) yields%
\begin{equation}
\sum_{i=1}^{N}\left(  \mathbf{I}+L_{0}\right)  \cdot\phi_{i}s_{i}+\left(
\mu-\mu_{\text{bif}}\right)  \sum_{i=1}^{N}L_{1}\cdot\phi_{i}s_{i}=-f_{\mu
}\left(  x;\mu\right)  \delta. \label{eqn:linear_v_expand}%
\end{equation}
The dot product on both sides of Eq. (\ref{eqn:linear_v_expand}) with
$\psi_{j}^{T}$ for $j=1,\ldots,N$ yields%
\begin{equation}
\left(  1+\lambda_{j}\right)  s_{j}+\left(  \mu-\mu_{\text{bif}}\right)
\sum_{i=1}^{N}\psi_{j}^{T}\cdot L_{1}\cdot\phi_{i}s_{i}=-\psi_{j}^{T}\cdot
f_{\mu}\left(  x;\mu\right)  \delta.
\end{equation}
Because $1+\lambda_{1}=0$, it follows that $s_{2}$, $\ldots$, $s_{N}$ are
negligible compared to $s_{1}$. To lowest order, solution of $s_{1}$ is
\begin{equation}
s_{1}=-\frac{\psi_{1}^{T}\cdot f_{\mu}\left(  x;\mu\right)  }{\left(  \mu
-\mu_{\text{bif}}\right)  \left(  \psi_{1}^{T}\cdot L_{1}\cdot\phi_{1}\right)
}\delta.
\end{equation}
Thus, an approximation of $v$ is
\begin{equation}
v=-\frac{\psi_{1}^{T}\cdot f_{\mu}\left(  x;\mu\right)  }{\left(  \mu
-\mu_{\text{bif}}\right)  \left(  \psi_{1}^{T}\cdot L_{1}\cdot\phi_{1}\right)
}\phi_{1}\delta.
\end{equation}
In steady state, the solution of the perturbed system alternates between two
states, $x_{\text{even}}$ and $x_{\text{odd}}$, where%
\begin{align}
x_{\text{even}}  &  =x_{\ast}+v,\\
x_{\text{odd}}  &  =x_{\ast}-v.
\end{align}
Recalling definition (\ref{eqn:gaindef}) of the gain yields
\begin{equation}
\text{$\Gamma$}=\left\vert \frac{v^{\left(  i\right)  }}{\delta}\right\vert
\sim\left\vert \frac{k}{\mu-\mu_{\text{bif}}}\right\vert ,
\end{equation}
where%
\begin{equation}
k=\frac{\psi_{1}^{T}\cdot f_{\mu}\left(  x_{\text{bif}};\mu_{\text{bif}%
}\right)  }{\psi_{1}^{T}\cdot L_{1}\cdot\phi_{1}}\phi_{1}^{\left(  i\right)
}. \label{eqn:k_def}%
\end{equation}

\subsection{An Example Illustrating the Limitation of the Linear Analysis}

As the first numerical example, we consider a one-dimensional cardiac map
\cite{hall02}:%

\begin{equation}
A_{n+1}=A_{\max}-\alpha\,e^{-D_{n}/\tau} \label{eqn:hall}%
\end{equation}
where $D_{n}=B-A_{n}$. Here, $A_{n}$ is the $n$th duration of action potential
of a cardiac cell under an electrical pacing with period $B$ (known as the
basic cycle length); $D_{n}$ is the $n$th diastolic interval. Using the
parameters $A_{\text{max}}=392.0$ ms, $\alpha=525.3$ ms, and $\tau=40.0$ ms,
Hall and Gauthier \cite{hall02} showed that a period-doubling bifurcation
occurs at $B=B_{\text{bif}}\approx455$ ms and the period-one fixed point of
the system is stable for $B>B_{\text{bif}}$. We refer interested readers to
\cite{hall02} for a bifurcation diagram exhibiting the period-doubling
bifurcation and other properties of this map.

\begin{figure}[tbh]
\centering
\includegraphics[width=6in]{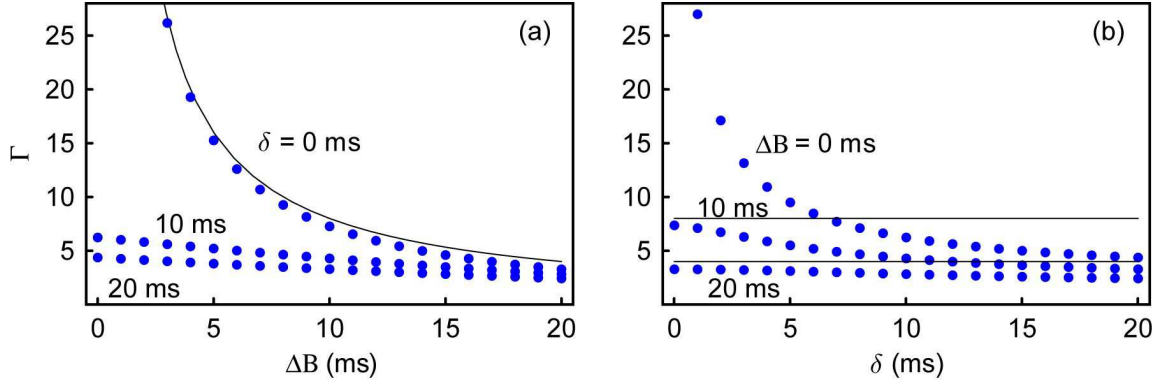} \caption{Variation of the gain
for the 1-D cardiac map as function of (a) $\Delta B=B-B_{\text{bif}}$ and (b)
perturbation amplitude $\delta$: the solid curves correspond to results from
the linear analysis and the dots correspond to numerical simulations. Note
that, in (b), the linear analysis predicts an infinitely large gain for
$\Delta B=0$ and any $\delta$.}%
\label{fig:hall_linear}%
\end{figure}

It is our interest to study the response of this cardiac map when the basic
cycle length $B$ is perturbed with an alternating perturbation in such a way
that $D_{n}=B+\left(  -1\right)  ^{n}\delta-A_{n}$. Following Theorem 1 and
evaluating the corresponding derivatives, one obtains the amplification gain
as%
\begin{equation}
\text{$\Gamma$}\sim\frac{2\tau}{B-B_{\text{bif}}}.
\end{equation}
Again, the theorem predicts i) the gain does not depend on perturbation
amplitude $\delta$ and ii) the gain approaches $\infty$ when the control
parameter $B\rightarrow B_{\text{bif}}$. For a given $B$, the magnitude of the
gain depends on the value of the parameter $\tau$. To verify these
observations, we compare results from theory and numerical simulations, as
shown in Fig. \ref{fig:hall_linear}. Here, Fig. \ref{fig:hall_linear} (a)
shows variation of the gain under changes in the control parameter $B$ when
the perturbation amplitude $\delta$ takes on various constant values and Fig.
\ref{fig:hall_linear} (b) shows variation of the gain under changes in
$\delta$ when $B$ takes on various constant values. In both figures, numerical
simulations indicate that the gain strongly depends on $\delta$. It is also
noted that the gain is not $\infty$ at $B=B_{\text{bif}}$ except when
$\delta=0$. These qualitative differences between the linear analysis and
numerical simulations therefore demand a higher-order nonlinear analysis of
the gain, which will be presented in the following section.
\begin{figure}[tbhtbh]
\centering
\includegraphics[width=6in]{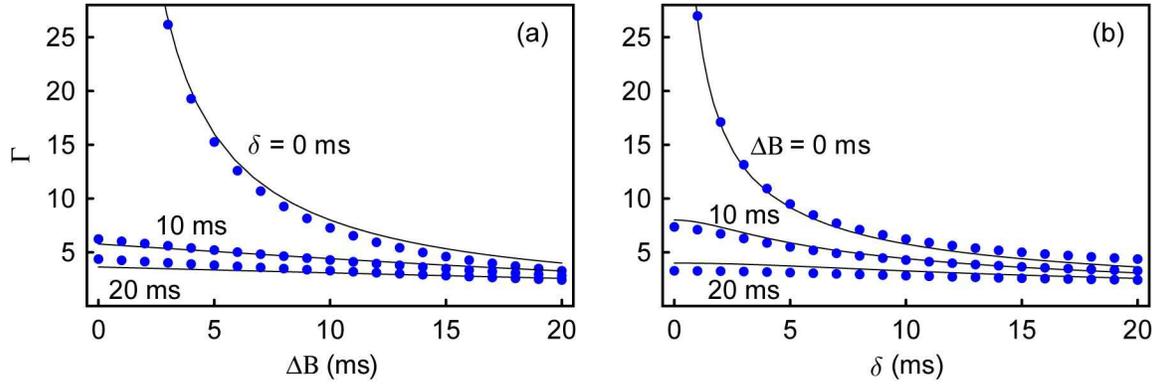}\caption{Variation of gain for
the 1-D cardiac map as function of (a) $\Delta B=B-B_{\text{bif}}$ and (b)
perturbation amplitude $\delta$: the solid curves correspond to results from
the higher-order nonlinear analysis and the dots correspond to numerical
simulations.}%
\label{fig:hall_nonlinear}%
\end{figure}

\section{Higher-Order Analysis}

Again, denote the fixed point of the map (\ref{eqn:generalMap}) at $\mu
=\mu_{\text{bif}}$ by $x_{\text{bif}}$. And again, denote by $\phi_{1}$ and
$\psi_{1}$ the right and left eigenvectors associated with the $-1$ eigenvalue
of the Jacobian $f_{x}\left(  x_{\text{bif}};\mu_{\text{bif}}\right)  $. Near
onset of the period-doubling bifurcation, the response of the perturbed map
(\ref{eqn:perturbedMap}) exhibits pre-bifurcation amplification. The following
theorem computes the amplification gain based on a higher-order nonlinear
analysis of the perturbed map.

\begin{thm2}
To leading order, the gain defined in Eq. (\ref{eqn:gaindef}) satisfies the
following relation,%
\begin{equation}
c\,\delta^{2}\,\Gamma^{3}+\left(  \mu-\mu_{\text{bif}}\right)  \Gamma
-\left\vert k\right\vert =0. \label{eqn:nonlineargain}%
\end{equation}
The coefficients in Eq. (\ref{eqn:nonlineargain}) are defined in the following subsection.
\end{thm2}

Here, again, $\phi_{1}^{\left(  i\right)  }$ represents the $i$th component of
$\phi_{1}$. For the special case of one-dimensional maps, the eigenvectors are
$\phi_{1}=\psi_{1}=1$ and thus%
\begin{align}
k  &  =\frac{f_{\mu}\left(  x_{\text{bif}};\mu_{\text{bif}}\right)  }%
{f_{xx}\left(  x_{\text{bif}};\mu_{\text{bif}}\right)  \cdot\frac{\partial
x_{\ast}}{\partial\mu}\left(  \mu_{\text{bif}}\right)  +f_{x\mu}\left(
x_{\text{bif}};\mu_{\text{bif}}\right)  }\\
c  &  =\frac{\left(  1/2f_{xx}\left(  x_{\text{bif}};\mu_{\text{bif}}\right)
\right)  ^{2}+1/6f_{xxx}\left(  x_{\text{bif}};\mu_{\text{bif}}\right)
}{f_{xx}\left(  x_{\text{bif}};\mu_{\text{bif}}\right)  \cdot\frac{\partial
x_{\ast}}{\partial\mu}\left(  \mu_{\text{bif}}\right)  +f_{x\mu}\left(
x_{\text{bif}};\mu_{\text{bif}}\right)  }.
\end{align}
Theorem 2 clearly indicates that the gain strongly depends on $\delta$ as well
as $\mu$. Especially at the bifurcation point, where $\mu=\mu_{\text{bif}}$,
the gain approaches to $\infty$ as a function $\delta^{-2/3}$. As shown by
numerical examples at the end of this section, the higher-order analysis
agrees well with numerical simulations.

\subsection{Proof of Theorem 2}

Heldstab \emph{et al.} \cite{heldstab83} seemingly were the first to conduct a
higher-order nonlinear analysis of the perturbed map (\ref{eqn:perturbedMap}).
Letting $\mu=\mu_{\text{bif}}$, they calculated the response of alternating
perturbations in one-dimensional maps. It was shown that deviation between
$x_{\text{even}}$ and $x_{\text{odd}}$ is proportional to $\delta^{1/3}$.
Following the same line of approach, we study the nonlinear response of maps
of any dimensions and consider the influence of both the control parameter
$\mu$ and the perturbation amplitude $\delta$. Naturally, the first attempt of
higher-order analysis is to include only quadratic terms, which, as shown in
the following, modifies the expressions for the median of the response as well
as modulates its amplitude. A retrospective examination reveals that the
contribution to the response from the cubic term is comparable to that from
the quadratic term and thus the cubic term has to be taken into account.

Again, we assume the perturbed response is a deviation $d_{n}$ away from the
unperturbed response $x_{\ast}$, defined in Eq. (\ref{eqn:md_fix}).
Substituting $x_{n}=x_{\ast}+d_{n}$ into Eq. (\ref{eqn:perturbedMap}) and
keeping the expansion of $f$ up to cubic order in $d_{n}$ yields%
\begin{equation}
d_{n+1}=L\cdot d_{n}+Q\cdot d_{n}\cdot d_{n}+C\cdot d_{n}\cdot d_{n}\cdot
d_{n}+\left(  -1\right)  ^{n}f_{\mu}\left(  x;\mu\right)  \delta+\ldots,
\label{eqn:md_expansion}%
\end{equation}
where%
\begin{align}
L  &  =f_{x}\left(  x;\mu\right) \nonumber\\
Q  &  =\frac{1}{2}f_{xx}\left(  x;\mu\right) \\
C  &  =\frac{1}{6}f_{xxx}\left(  x;\mu\right)  .
\end{align}
To account for the alternating feature of the perturbation, we decompose the
steady state of \textbf{$d$}$_{n}$ into an oscillating component $\left(
-1\right)  ^{n}v$ and a mean value $m$. Substituting \textbf{$d$}$_{n}=\left(
-1\right)  ^{n}v+m$ into Eq. (\ref{eqn:md_expansion}) yields%
\begin{align}
m  &  =L\cdot m+Q\cdot v\cdot v\label{eqn:steady_m}\\
v  &  =-L\cdot v-2Q\cdot m\cdot v\nonumber\\
&  -C\cdot v\cdot v\cdot v-f_{\mu}\left(  x;\mu\right)  \delta.
\label{eqn:md_vmap_high}%
\end{align}

Because $\mu$ is close to $\mu_{\text{bif}}$, $L$ can be approximated as
\begin{equation}
L=L_{0}+\left(  \mu-\mu_{\text{bif}}\right)  L_{1},
\end{equation}
where $L_{0}$ and $L_{1}$ are defined in Eqs. (\ref{eqn:md_L0}) and
(\ref{eqn:md_L1}). Again, denote the eigenvalues of $L_{0}$ by $-1=\lambda
_{1}<\lambda_{2}\leq\ldots\leq\lambda_{N}<1$ and denote the corresponding
right and left eigenvectors by $\phi_{1},\phi_{2},\ldots,\phi_{N}$ and
$\psi_{1}^{T},\psi_{2}^{T},\ldots,\psi_{N}^{T}$, respectively. We assume the
eigenvectors are normalized such that $\psi_{i}^{T}\cdot\phi_{j}=1$ if $i=j$
and $\psi_{i}^{T}\cdot\phi_{j}=0$ if $i\neq j$.

It follows from Eq. (\ref{eqn:steady_m}) that, in steady state,%
\begin{equation}
m=\tilde{Q}\cdot v\cdot v,
\end{equation}
where%
\begin{equation}
\tilde{Q}=\left(  \mathbf{I}-L_{0}\right)  ^{-1}\cdot Q.
\end{equation}
To solve for the steady-state solution of $v$, we again decompose $v$ as%
\begin{equation}
v=\sum_{i=1}^{N}s_{i}\phi_{i}=\Phi\cdot s, \label{eqn:v_phi_s}%
\end{equation}
where $\Phi=\left(  \phi_{1},\ldots,\phi_{N}\right)  $ and $s=\left(
s_{1},\ldots s_{N}\right)  ^{T}$. Substituting Eq. (\ref{eqn:v_phi_s}) into
Eq. (\ref{eqn:md_vmap_high}) and taking the dot-product of this expression
with $\psi_{j}$ for $j=1,\ldots,N$ on both sides yields%
\begin{align}
&  \left(  1+\lambda_{j}\right)  s_{j}+\left(  \mu-\mu_{\text{bif}}\right)
\sum_{i=1}^{N}\psi_{j}^{T}\cdot L_{1}\cdot\phi_{i}s_{i}\\
&  =-2\psi_{j}^{T}\cdot Q\cdot\left(  \tilde{Q}\cdot\left(  \Phi\cdot
s\right)  \cdot\left(  \Phi\cdot s\right)  \right)  \cdot\left(  \Phi\cdot
s\right) \\
&  -\psi_{j}^{T}\cdot C\cdot\left(  \Phi\cdot s\right)  \cdot\left(  \Phi\cdot
s\right)  \cdot\left(  \Phi\cdot s\right)  -\psi_{j}^{T}\cdot f_{\mu}\left(
x;\mu\right)  \delta.
\end{align}
Because $1+\lambda_{1}=0$, it follows that $s_{2}$, $\ldots$, $s_{N}$ are
higher order compared to $s_{1}$ and thus $v\approx\phi_{1}s_{1}$. Then, the
governing equation for $s_{1}$ is, to lowest order,%
\begin{equation}
\left(  \bar{Q}+\bar{C}\right)  s_{1}^{3}+\left(  \mu-\mu_{\text{bif}}\right)
\bar{L}s_{1}+\psi_{1}^{T}\cdot f_{\mu}\left(  x;\mu\right)  \delta=0,
\label{eqn:s1}%
\end{equation}
where%
\begin{align}
\bar{L}  &  =\psi_{1}^{T}\cdot L_{1}\cdot\phi_{i}\\
\bar{Q}  &  =2\psi_{1}^{T}\cdot Q\cdot\left(  \tilde{Q}\cdot\phi_{1}\cdot
\phi_{1}\right)  \cdot\phi_{1}\\
\bar{C}  &  =\psi_{1}^{T}\cdot C\cdot\phi_{1}\cdot\phi_{1}\cdot\phi_{1}.
\end{align}

The perturbed response governed by Eq. (\ref{eqn:s1}) is valid for both
supercritical and subcritical period-doubling bifurcations. However, for a
supercritical period-doubling bifurcation such as the one mediating cardiac
alternans, $s_{1}=0$ is the only solution when $\delta=0$ and $\mu
>\mu_{\text{bif}}$; that is to say, before the occurrence of a supercritical
period-doubling bifurcation, the only solution of the unperturbed map
(\ref{eqn:generalMap}) is the period-one solution. Thus, for a supercritical
period-doubling bifurcation, it follows from Eq. (\ref{eqn:s1}) that
\begin{equation}
\left(  \bar{Q}+\bar{C}\right)  \bar{L}>0.
\end{equation}
In the following, we restrict our attention to supercritical period-doubling
bifurcations to simplify the result.

In steady state, the solution of the perturbed system alternates between two
states, $x_{\text{even}}$ and $x_{\text{odd}}$, where%
\begin{align}
x_{\text{even}}  &  =x_{\ast}+v.\\
x_{\text{odd}}  &  =x_{\ast}-v.
\end{align}
Recalling the definition of the gain yields
\begin{equation}
\text{$\Gamma$}=\left\vert \frac{\phi_{1}^{\left(  i\right)  }s_{1}}{\delta
}\right\vert . \label{eqn:gain_define_tmp}%
\end{equation}
Substituting Eq. (\ref{eqn:gain_define_tmp}) into (\ref{eqn:s1}) yields
\begin{equation}
c\,\delta^{2}\,\Gamma^{3}+\left(  \mu-\mu_{\text{bif}}\right)  \Gamma
-\left\vert k\right\vert =0, \label{eqn:gamma_eqn}%
\end{equation}
where%
\begin{equation}
c=\frac{\bar{Q}+\bar{C}}{\bar{L}\left(  \phi_{1}^{\left(  i\right)  }\right)
^{2}},
\end{equation}
and $k$ is defined in Eq. (\ref{eqn:k_def}).

\subsection{Examples Applying the Nonlinear Analysis}

\subsubsection{One-Dimensional Cardiac Model}

\begin{figure}[tbh]
\centering
\includegraphics[width=4in]{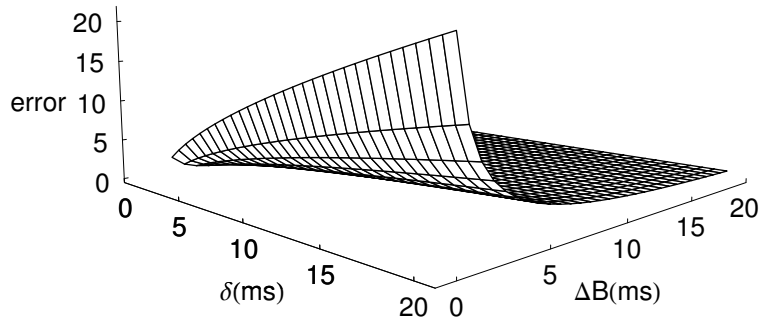}\caption{Discrepancy between the linear
and nonlinear analyses for the the 1-D cardiac map. Here, the error is defined
as $\left(  \Gamma_{\text{L}}-\Gamma_{\text{N}}\right)  /\Gamma_{\text{N}}$,
where $\Gamma_{\text{L}}$ is the result from the linear analysis and
$\Gamma_{\text{N}}$ is the result from the nonlinear analysis.}%
\label{fig:error}%
\end{figure}

Using the higher-order analysis, we reexamine the one-dimensional cardiac
model described by Eq. (\ref{eqn:hall}). Results from the higher-order
analysis agree well with numerical simulations, as seen in Fig.
\ref{fig:hall_nonlinear}. It is observed from Eq. (\ref{eqn:gamma_eqn}), as
well as the numerical results, that the gain has a strong dependence on
$\delta$ and the gain in general is not $\infty$ at $B=B_{\text{bif}}$ except
when $\delta=0$. The discrepancy between the linear and nonlinear analyses can
be seen clearer in Fig. \ref{fig:error}, which shows the error between the
linear and nonlinear predictions in two-dimensional parameter space $\left(
\Delta B,\delta\right)  $. The figure shows that the error increases as
$\delta$ increases and for the same $\delta$ the error decreases as $\Delta B$
increases. 

\subsubsection{Two-Dimensional Cardiac Model}

\begin{figure}[tbh]
\centering
\includegraphics[width=6in]{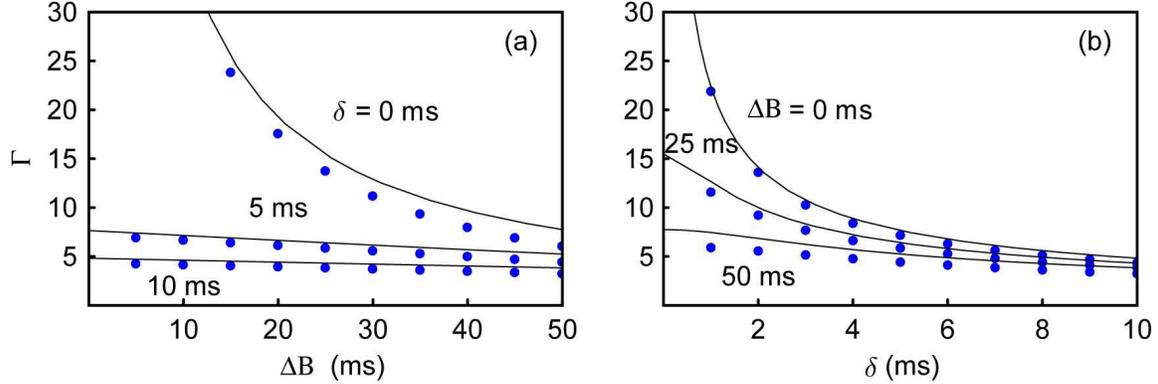} \caption{Variation of the gain for the
two-dimensional cardiac map as function of (a) $\Delta B=B-B_{\text{bif}}$ and
(b) perturbation amplitude $\delta$: the solid curves correspond to results
from the higher-order nonlinear analysis and the dots correspond to numerical
simulations.}%
\label{fig:jalife}%
\end{figure}

We now consider a two-dimensional map presented by Chialvo \emph{et al.
}\cite{chialvo90} and later studied by Hall and Gauthier \cite{hall02}%
\begin{align}
A_{n+1}  &  =(1-M_{n+1})\left(  A_{\max}-\alpha\,e^{-D_{n}/\tau}\right)
,\label{J}\\
M_{n+1}  &  =[1-(1-M_{n})e^{-A_{n}/\tau_{2}}]e^{-D_{n}/\tau_{2}}, \label{JM}%
\end{align}
where $D_{n}=B-A_{n}$, $A_{n}$ is the $n$th duration of action potential of a
cardiac cell, $B$ is the basic cycle length representing the period of applied
electrical pacing, and $D_{n}$ is the $n$th diastolic interval. Using the
parameters $A_{\max}=490.9$ ms, $\alpha=569.0$ ms, $\tau_{1}=64.0$ ms,
$\tau_{2}=38.0$ ms, $D_{\min}=38.0$ ms, Hall and Gauthier showed that a
transition to alternans occurs at $B=B_{\text{bif}}\approx455$ ms
\cite{hall02}. We refer interested readers to \cite{hall02} for a bifurcation
diagram exhibiting the period-doubling bifurcation and other properties of
this map. We study the response of this map under alternating perturbations.
Figure \ref{fig:jalife} shows the variation of the amplification gain under
changes in $B$ and $\delta$. Again, it is seen that good agreement is achieved
between the higher-order analysis and the numerical
simulations.

\section{Conclusion and Discussion}

Using both linear and higher-order analyses, we have investigated the
pre-bifurcation amplification near a period-doubling bifurcation in
multi-dimensional iterated maps. Through numerical examples, it has been
estabilished that the qualitative features of the amplification gain can only
be captured using the higher-order analysis. In general, the gain is a
nonlinear function of $\mu$ and $\delta$ as seen from Eq. (\ref{eqn:gamma_eqn}%
). In the $\left(  \mu,\delta\right)  $ parameter region, the gain is finite
everywhere except at $\left(  \mu_{\text{bif}},0\right)  $. Moreover, the rate
of divergence as $\left(  \mu,\delta\right)  $ approaches to this singular
point depends on the path taken. Under two situations, this nonlinear function
can be significantly simplified. First, when $\left\vert \mu-\mu_{\text{bif}%
}\right\vert \gg\delta^{2/3}$, the first term in Eq. (\ref{eqn:gamma_eqn})
becomes negligible compared to the others, resulting in the simplified
expression%
\begin{equation}
\left(  \mu-\mu_{\text{bif}}\right)  \Gamma-\left\vert k\right\vert =0.
\end{equation}
It follows that the gain can be approximated by
\begin{equation}
\Gamma=\left\vert \frac{k}{\mu-\mu_{\text{bif}}}\right\vert ,
\label{eqn:linearGain}%
\end{equation}
which is the result of linear analysis obtained in Section 2.
\begin{figure}[tbh]
\centering
\includegraphics[width=3in]{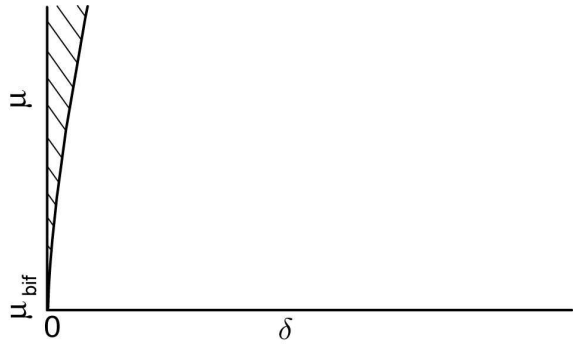}\caption{Range of validity: the
linear analysis is valid in the hatched area; the higher-order analysis is
valid in the whole region. Here, it has been assumed that region of interest
is when $\mu>\mu_{\text{bif}}$. }%
\label{fig:smoothrov}%
\end{figure}On the other hand, when $\left\vert \mu-\mu_{\text{bif}%
}\right\vert \ll\delta^{2/3}$, the second term in Eq. (\ref{eqn:gamma_eqn})
becomes negligible compared to the others, resulting in the simplified
expression%
\begin{equation}
c\,\delta^{2}\,\Gamma^{3}-\left\vert k\right\vert =0.
\end{equation}
It follows that the gain can be approximated by
\begin{equation}
\Gamma=\sqrt[3]{\left\vert \frac{k}{c}\right\vert }\delta^{-2/3}.
\label{eqn:2thirdGain}%
\end{equation}
A special case of the latter situation is when $\mu=\mu_{\text{bif}}$, which
was studied by Heldstab \emph{et al.} using one-dimensional maps
\cite{heldstab83}. Note that Heldstab \emph{et al.} did not include the
contribution due to the cubic order expansion of $v$ (\emph{i.e.} the
contribution of $C$), although their result is still valid for the logistic
map because $C=0$ for this map. Our analysis allows us to predict the range of
validity for the linear and the higher-order analyses, as shown in Fig.
\ref{fig:smoothrov}. Especially, the linear analysis is only valid in a
restricted area, which shrinks to zero as $\mu$ approaches $\mu_{\text{bif}}$,
cf. the numerical illustration in Fig. \ref{fig:error}.

\begin{figure}[tbh]
\centering
\includegraphics[width=6in]{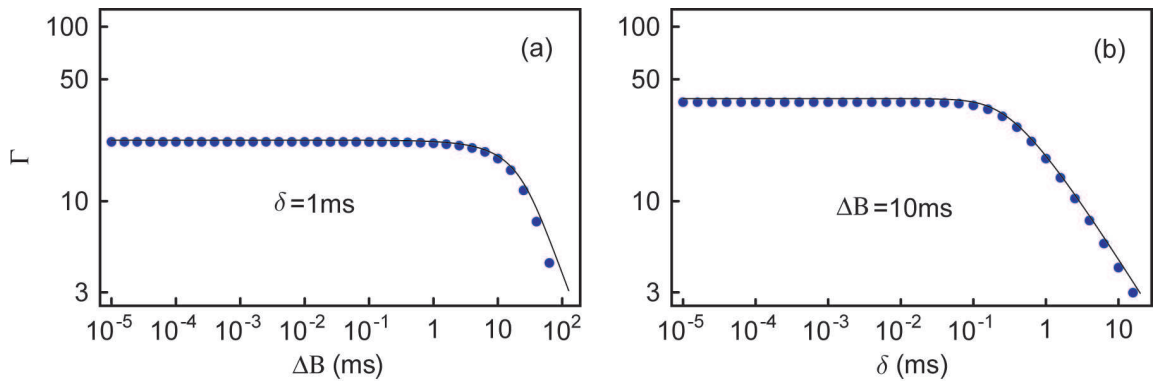}\caption{Saturation of the
amplification gain in the two-dimensional cardiac map illustrated by (a)
$\log\Gamma$ vs. $\log\Delta B$, where $\Delta B=B-B_{\text{bif}}$ and (b)
$\log\Gamma$ vs. $\log a$: the solid curves correspond to results from the
higher-order analysis and the dots correspond to numerical simulations.}%
\label{fig:saturation}%
\end{figure}

The two aforementioned special situations and the analysis of the range of
validity lead to an interesting observation on the relationship between the
gain $\Gamma$ and the parameters. First, when $\delta$ takes on a constant
value, one can see from Figure \ref{fig:saturation} (a) that $\log\Gamma$ is
almost proportional to $\log\left(  \mu-\mu_{\text{bif}}\right)  ^{-1}$ for
large $\mu$, and $\log\Gamma$ increases and saturates to a constant as $\mu$
decreases to a sufficiently small value. Conversely, when $\mu$ takes on a
constant value, one can see from Figure \ref{fig:saturation} (b) that
$\log\Gamma$ is almost proportional to $\log\delta^{-2/3}$ for large $\delta$,
and $\log\Gamma$ increases and saturates to a constant as $\delta$ decreases
to a sufficiently small value. We note that a similar saturation effect in
pre-bifurcation amplification was previously observed by Kravtsov and
Surovyatkina \cite{kravtsov03} and Surovyatkina \cite{surovyatkina04}, who
considered random noise in a one-dimensional map as $\mu$ is varied to
approach $\mu_{\text{bif}}$. Although the saturation of the gain is a
universal property of dynamical systems, we note that it is very difficult to
observe experimentally such phenomenon on paced cardiac tissues because of the
existence of noise and the limitation on measurement resolution.

\begin{figure}[tbh]
\centering
\includegraphics[width=3in]{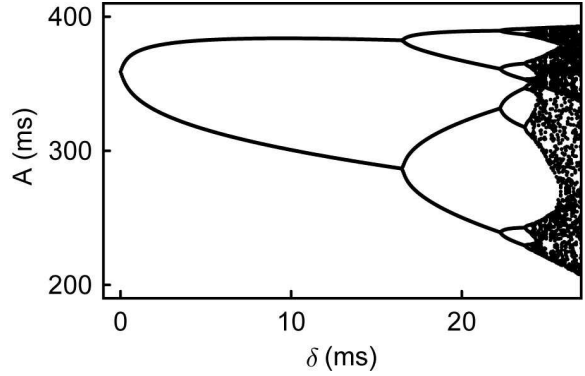}\caption{Variation of perturbed response
under changes in perturbation amplitude $\delta$ for the two-dimensional
cardiac map described in Eqs. (\ref{J}) and (\ref{JM}). Here, the control
parameter is chosen to be $B=B_{\text{bif}}+10$ ms.}%
\label{fig:chaos}%
\end{figure}

Because pre-bifurcation amplification is a property of the period-doubling
bifurcation, its observable region in terms of the control parameter and the
perturbation amplitude is problem-dependent and limited in parameter space to
a vicinity of the bifurcation point. Under some circumstances,
de-amplification gains (gains less than $1$) may be obtained for a control
parameter $\mu$ sufficiently far away from $\mu_{\text{bif}}$. Moreover, other
responses rather than the presumed period-two solutions may also exist. For
example, as shown in Fig. \ref{fig:chaos}, a period-doubling cascade to chaos
is observed in the two-dimensional cardiac model under increase of the
perturbation amplitude $\delta$. These observations suggest that caution
should be taken when using pre-bifurcation amplification as a technique to
detect the existence of period-doubling bifurcations. 

\begin{center}
\textbf{Acknowledgments}
\end{center}

Support of the National Institutes of Health under grant 1R01-HL-72831 and the
National Science Foundation under grants DMS-9983320 and PHY-0243584 is
gratefully acknowledged.

\end{document}